\documentclass[11pt,showpacs]{revtex4}
\usepackage{graphicx}
\begin{document}

\preprint{}

\title{Gravitational instanton, inflation and cosmological constant}

\author{She-Sheng Xue}

\email{xue@icra.it}

\affiliation{ICRA and
Physics Department, University of Rome ``La Sapienza", 00185 Rome, Italy}



\begin{abstract}
Quantum fluctuation of unstable modes about gravitational instantons causes 
the instability of flat space at finite temperature, leading to the 
spontaneous process of nucleating quantum black holes. The density of vacuum 
energy-gain in such process gives the cosmological term in the Einstein equation. 
This naturally results in the inflationary phase of Early Universe. While the 
reheating phase is attributed to the Hawking radiation of these quantum 
black holes. In the Standard cosmology era, this cosmological term depends on the reheating
temperature and asymptotically approaches to the cosmological constant 
in matter domination phase, consistently with current observations.    
\end{abstract}


\maketitle

\vskip0.2cm
\noindent{\it Introduction.}\hskip0.3cm 
In recent years, as observational data concerning on cosmology are rapidly accurated, 
the theoretical understanding of our Universe has been greatly profounded ever before. 
The inflation\cite{linde} in early Universe and the acceleration of present 
Universe\cite{wein} are two most important issues in modern cosmology and fundamental 
physics. Both issues are closely related to the vacuum-energy represented by the 
cosmological constant in Einstein equation.
Much effort in understanding these issues has been made for many decades 
and there are many interesting ideas and innovative theoretical developments based 
on either simple models\cite{q} or complex theories\cite{s}. In these approaches, 
a scalar field slowly-rolling downwards an effective potential, 
which mimics a possible vacuum-energy variation, plays a crucial r\^ole in driving 
inflation; whereas this scalar field with a mass of order the current Hubble scale, can 
possibly account for acceleration. We attempt to study
these two issues within a framework based on the vacuum-energy variation in 
spacetime decay due to an unstable quantum fluctuation about 
gravitational instantons.
      
\vskip0.2cm
\noindent{\it Instability of flat space at finite temperature.}\hskip0.3cm
The attractive nature of gravity that cannot be screened is the essential reason for
many inevitable instabilities of classical gravitating systems. One might worry about
flat-space metastable state and the stability of flat space against quantum-field 
tunneling. The positive-energy theorem {\it \'a la} Schoen and Yau\cite{yao} shows that
the total energy of asymptotically flat manifolds is positively semidefinite and only
Minkowski space has zero energy. This
precludes the possibility of flat space at zero temperature decaying by any mechanism. 

In Ref.\cite{gross}, authors explored two 
instabilities of flat space at finite temperature $T$, which is attributed to a 
thermal equilibrium of gravitons and other particles. For large-wavelength density 
fluctuations of the thermal gravitons,
a Jeans instability is expected to occur. Another source of instability is the nucleation 
of quantum black holes, due to an effect of quantum fluctuations about a gravitational instanton, 
which is the the Euclidean section of the Schwarzschild metric $g_{ab}^S$. The classical Euclidean 
action of $N$ non-interacting instantons is given by, 
\begin{equation}
I_i^n=NI_i,\hskip0.5cm I_i={1\over2}\beta M,\hskip0.5cm \beta=(8\pi GM),
\label{inactionc}
\end{equation}
where $I_i$ is the classical Euclidean action for an instanton, $M$ is not a fixed mass, 
but rather determined in 
terms of the temperature $T=\beta^{-1}$. The inverse temperature $\beta$ is the 
period in Euclidean time. 

About this instanton saddle point $g_{ab}^S$, 
in addition to quantum fluctuations of stable
modes of thermal gravitons, there is an unstable (negative) mode 
($\epsilon \tilde \phi_{ab}$)\cite{gross,unstable}, namely $g_{ab}=g_{ab}^S+\epsilon \tilde \phi_{ab}$. Its
quantum fluctuation decreases the classical action (\ref{inactionc}) by 
$\delta I_i=- 9.4\cdot 10^{-4}\epsilon^2 /(GM)^2$. Its contributions to the partition functional integral 
give rise to an imaginary part in the free energy (effective action). In Ref.\cite{gross},
this is interpreted as a finite lifetime for decay of flat space at finite temperature $T$. 
The decay proceeds by quantum fluctuations spontaneously nucleating quantum black holes of 
radius $R=(4\pi T)^{-1}$ and mass $M=(8\pi GT)^{-1}$, where the Planck mass $m_p=G^{-1/2}$. 
Assuming the configuration of 
non-interacting instantons, one calculated the rate (per 
unit volume $\delta V$) for nucleating these quantum black holes\cite{gross,rate}
\begin{eqnarray}
\Gamma(T)\! &\equiv& \!{\delta^2 N\over\delta t\delta V}\!= \!0.87 T ({m_p\over T})^\theta {m_p^3\over 64 \pi^3}
\exp\Big[\!-\!{m_p^2\over 16 \pi T^2}\Big],
\label{inrate}\\
\theta &=&{1\over 45}(212n_2\!-\!{233\over 4}n_{3/2}\!-\!13n_1\!+\!{7\over 4}n_{1/2}\!+\!n_0),
\label{theta}
\end{eqnarray}
where $n_s$ is the number of massless spin-$s$ fields. 
At high-temperature $T\sim m_p$, the rate (\ref{inrate}) of the nucleation 
is maximum and exponentially suppressed at low-temperature $T\ll m_p$.

This shows that (i) stable and unstable quantum fluctuations of gravitational field 
contribute to the vacuum energy and (ii) the vacuum of flat space at finite 
temperature $T$ is energetically unstable and decays by spontaneously producing 
quantum black holes. The energy-gain in this vacuum-decay should have an important 
impact on the evolution of early Universe and be related to the cosmological constant 
of present Universe. 

\vskip0.2cm
\noindent{\it Black hole nucleation and cosmological term.}\hskip0.3cm
Based on Eq.(\ref{inrate}), the total number $N$ of quantum black holes 
nucleated is given by
\begin{equation}
N = \int (-g)^{1/2}d^4x \Gamma(T).  
\label{number}
\end{equation}
We assume that (i) these quantum black holes are free from interacting each other and 
their spatial distribution is homogeneous; (ii) the kinetic energy of these quantum 
black holes is much smaller than their mass-energy $M$.
Analogously to Eq.(\ref{inactionc}), the effective Euclidean action of these
non-interacting black holes is given by using Eq.(\ref{number}),
\begin{equation}
I_{\rm BH}=\int (-g)^{1/2}d^4x \Gamma(T){1\over2}\beta M.  
\label{eaction}
\end{equation}
The corresponding effective action in Minkowski time $t$ is obtained by 
the Wick rotation $\beta=\int d\beta = -i\int dt$, 
\begin{equation}
S_{\rm BH} = \int (-g)^{1/2}d^4x \rho_\Lambda(t),  
\label{inactionq}
\end{equation}
where
\begin{eqnarray}
\rho_\Lambda(t)&=&{1\over2}\int_{t_0}^t d\tau M(\tau)\Gamma [T(\tau)],
\nonumber\\
&=& {0.87m_p^4\over 1024\pi^4}\int_{t_0}^td\tau 
\left({1\over T(\tau)}\right)^\theta e^{-{1\over 16\pi T^2(\tau)}} 
\label{rho}
\end{eqnarray}
which is the energy-density of quantum black holes nucleated from initial time $t_0$ 
and temperature $T_0$ to final time $t$ and temperature $T(t)$. The energy-gain 
$\delta E_\Lambda=\rho_\Lambda\delta V$ in the spontaneous nucleation process. In 
Eq.(\ref{rho}) and henceforth, the temperature $T$ and time $t$ are in Planck units.

Armed with the effective action (\ref{inactionq}) together with actions for gravity 
and thermal particles, we arrive at the Einstein equation
\begin{equation}
R_{ab}- {1\over2}g_{ab} R + g_{ab}\Lambda = 8\pi G T_{ab},
\label{equation} 
\end{equation}
where the cosmological term $\Lambda =8\pi G \rho_\Lambda$  
is originated from the effective action (\ref{inactionq}). 
Such cosmological term has two features: (i) its geometric origin from gravitational 
instantons, appearing in the l.h.s. of Eq.(\ref{equation}); (ii) its energetic 
origin from the energy-gain $\delta E_\Lambda$ of nucleation of quantum black holes 
$\delta E_\Lambda=-\delta E$, where $\delta E<0$ is the vacuum-energy variation of 
the flat space at finite temperature. 
The second point gives negative pressure $p_\Lambda\equiv -\delta E_\Lambda/\delta V 
=-\rho_\Lambda$. The corresponding energy-momentum tensor
$T^\Lambda_{ab}=-g_{ab}\rho_\Lambda$. The total energy conservation
$(T_{\mu\nu}+T^\Lambda_{\mu\nu})^{;\mu}=0$.

\vskip0.2cm
\noindent{\it Inflation in early Universe.}\hskip0.3cm
We consider the post-Planckian era of early Universe, for the time $t\geq t_0=1$. 
The flat space is at the temperature $T\leq T_0=1$. The spontaneously energy-gain 
process of nucleating quantum black holes is bound to occur.
Suppose that the Universe is given by the Robertson-Walker line element with the 
scale factor $a(t)$ and zero curvature $k=0$, then the Einstein equation 
(\ref{equation}) describing Universe expansion becomes,
\begin{equation}
H^2\equiv \left({\dot a\over a}\right)^2 = {8\pi G \over 3}(\rho_m(t) + \rho_\Lambda(t)).
\label{requation} 
\end{equation}
where $\rho_m$ is the energy-densities of thermal particles and $\rho_\Lambda(t)$ is given by
Eq.(\ref{rho}) with the initial time $t_0=1$ and temperature $T_0=1$.
The Universe expands and its temperature $T(t)$ decreases, $a(t)T(t)=$constant for 
the entropy-conservation. As results, the energy-densities $\rho_m(t)$
decreases, whereas $\rho_\Lambda(t)$ (\ref{rho}) 
asymptotically approaches a constant $\bar\rho_\Lambda$ for $t\gg 1$. 
This implies an inflationary Universe. How does such inflation end? 

\vskip0.2cm
\noindent{\it Hawking radiation and reheating.}\hskip0.3cm
When the Universe temperature $T(t)$ is smaller than the temperature 
$T(\tau)=1/(8\pi GM(\tau))$ of quantum black holes that are created at an 
earlier time $\tau <t$, these quantum black holes loss their masses by the 
Hawking radiation. On the other hand, accretion occurs if $T(t)>T(\tau)$. 
If we were only to consider emission and absorption of gravitons, 
the mass-variation of quantum black holes is given by\cite{hawking,accretion} 
\begin{equation}
{\delta M(\tau)\over \delta\tau} = {\pi^2\over 15}[T^4(t)-T^4(\tau)]4\pi R^2(\tau), 
\label{hrate1} 
\end{equation}
where the black hole size $R(\tau)=2M(\tau)$. For $T(\tau)\gg T(t)$, we approximately
obtain,
\begin{equation}
M_H(t)\simeq M(\tau)[1-{2\pi^2\over 5}T^3(\tau)(t-\tau)]^{1/3}, 
\label{hrate2} 
\end{equation}
indicating that quantum black holes loss mass. We speculate that the extremal minimum of black hole
mass is the order of the Planck scale, where Eqs.(\ref{hrate1},\ref{hrate2}) are not applicable.   
 
The Hawking process (\ref{hrate2}) occurs, contemporaneously with quantum black hole nucleation. Due to the 
Hawking radiation, the energy-density $\rho_\Lambda(t)$ (\ref{rho}) 
of quantum black holes is reduced by
\begin{eqnarray}
\Delta\rho_\Lambda(t) &=&{1\over2}\!\int_{t_0}^t d\tau \Delta M\Gamma [T(\tau)],
\label{deltarho}\\
&=& {0.87m_p^4\over 128\pi^3}\int_{t_0}^td\tau \Delta M
\left({1\over T(\tau)}\right)^{(\theta -1)}e^{-{1\over 16\pi T^2(\tau)}},\nonumber
\end{eqnarray}
where $\Delta M\!=\!M_H(t)\!-\!M(\tau)$ and $M_H(t)$ is given by Eq.(\ref{hrate2}). 
The energy-density $\rho_\Lambda(t)$ in Eq.(\ref{requation}) should be replaced by 
\begin{equation}
\rho(t)=\rho_\Lambda(t)+\Delta\rho_\Lambda(t)<\rho_\Lambda(t).
\label{totalrho}
\end{equation}
This shows that the energy-gain $\rho_\Lambda(t)$ in quantum black hole nucleation is 
converted into the radiation energy $\Delta\rho_\Lambda(t)$, which reheats the Universe.

Assuming $a(t)T(t)\simeq 1$ and $\theta=212/45$(graviton), 
$t_0\simeq 1$ and $T_0\simeq 1$, we numerically integrate Eqs.(\ref{requation},\ref{totalrho}) 
for the evolution of Universe, by taking into account 
both the nucleation process (\ref{inrate},\ref{rho}) and the Hawking process (\ref{hrate2},\ref{deltarho}). 
At the beginning ($t< O(10^2)$), the nucleation process is dominate over the Hawking process.
The variation of energy-density $\rho(t)$ (\ref{totalrho}) is rather slow, so that the 
``graceful exit'' problem is avoided and the evolution of Universe is inflationary. 
At the end ($t> O(10^2)$), however, the Hawking process eventually ends 
the inflation and reheats Universe.

\vskip0.2cm
\noindent{\it Numerical results.}\hskip0.3cm       
In the initial phase $t\sim 1$, the energy-density of quantum black holes 
$\rho_\Lambda(t)$ (\ref{rho}) is negligible, compared with the energy-density of
thermal graviton gas $\rho_m={\pi^2\over 15} T^4$. The solution to Eq.(\ref{requation}) is radiative,
\begin{equation}
a_1= a_0 t^{1/2},\hskip0.3cm T_1=T_0 t^{-1/2},
\label{radiative} 
\end{equation} 
where $a_0\sim 1$ and $T_0\sim 1$ are initial scaling-factor and temperature.
$a_0T_0\sim 1$ implies that the initial entropy $S_0=(a_0T_0)^3$ is given 
by $O(1)$ quantum states of Planck energy in the Planck volume.
As the time $t$ increases, $\rho_m(t)$ decreases, $\rho_\Lambda(t)$ increases 
and becomes dominant in Eq.(\ref{requation}). 

\begin{figure}[th]
\begin{center}
\includegraphics[]{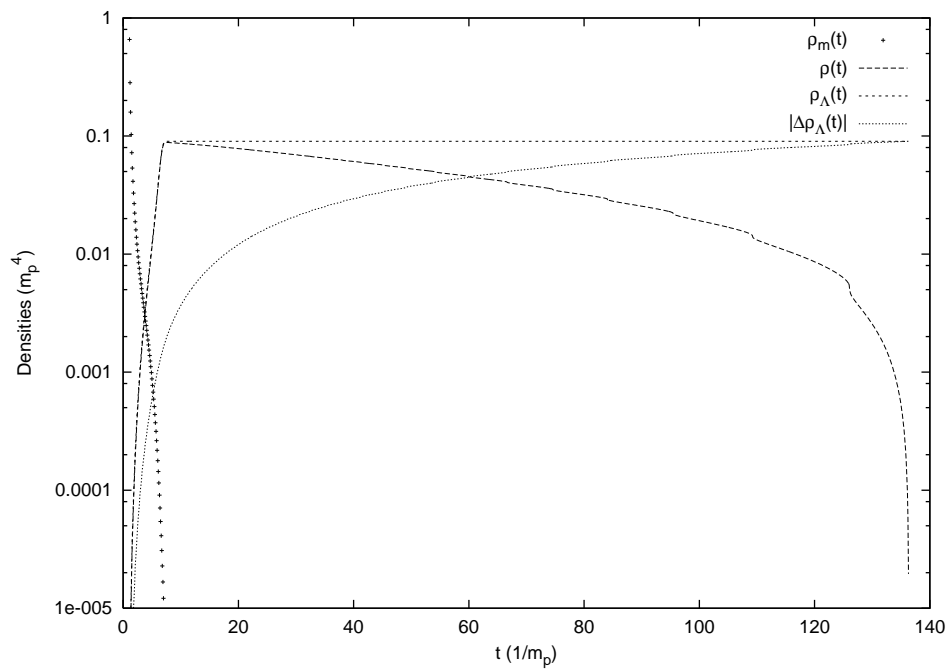}
\end{center}
\caption{The energy-densities $\rho_m(t)$, $\rho_\Lambda(t)$, $\rho(t)$ and 
$|\Delta\rho_\Lambda(t)|$ as functions of time.}%
\label{densities}%
\end{figure}

\begin{figure}[th]
\begin{center}
\includegraphics[]{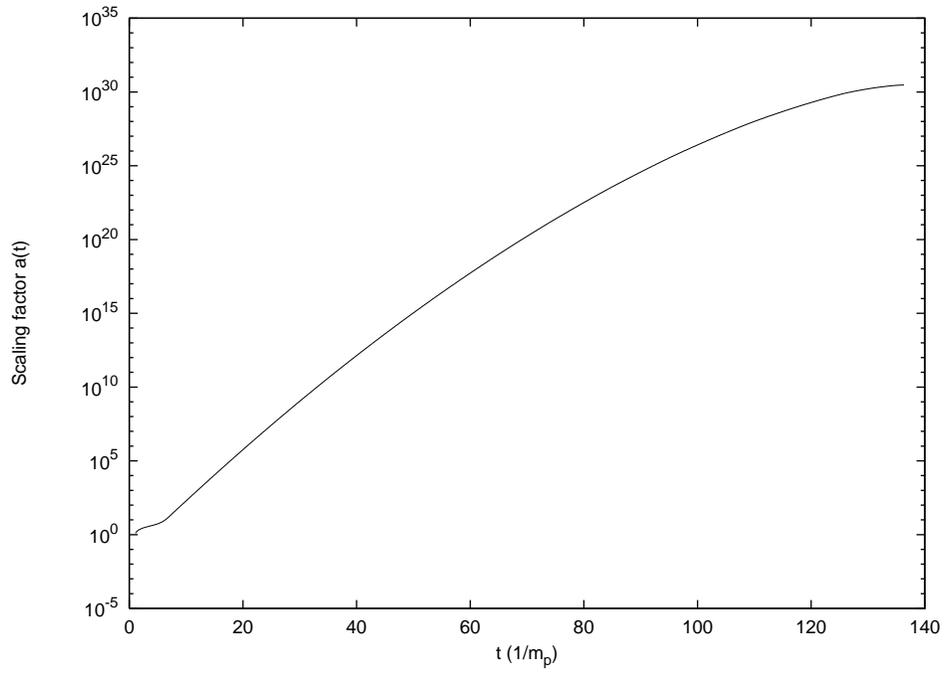}
\end{center}
\caption{The scaling factor $a(t)$ and temperature $T(t)=1/a(t)$ as functions of time.}%
\label{inflation}%
\end{figure}

In Fig.[\ref{densities}], we plot $\rho_m(t)$ (cross line) 
and $\rho_\Lambda(t)$ (short-dash line) in terms of the time $t$.
It is shown that (i) the pre-inflationary phase (\ref{radiative}) for $t<10$; (ii) the inflationary phase 
for $t>10$, where $\rho_m$ is vanishing and $\rho_\Lambda(t)$ is approaching to an 
asymptotic value $\bar \rho_\Lambda\simeq 9.03\cdot 10^{-2}m_p^4$. In Fig.[\ref{densities}], 
we also plot the radiation energy $|\Delta\rho_\Lambda(t)|$ (\ref{deltarho}) (dot-line) 
and the energy-density $\rho(t)$ 
(\ref{totalrho}) (long-dash line), showing that $\rho(t)$ slowly varies for $8<t<100$ and 
$\rho(t)\rightarrow 0$ for $t>100$. Correspondingly, in Fig.[\ref{inflation}], we plot $a(t)$ 
and $T(t)\simeq 1/a(t)$, which show that (i) the pre-inflationary phase (\ref{radiative}) for $t<4$, 
(ii) an exponential inflation $a(t)\simeq a_0\exp (N_et)$ for $4<t<110$ and (iii) $a(t)$
approaching $10^{30}$ for $t>110$. 

We might consider that the inflation ends at $t\simeq t_{h}$, when the quantum 
black hole mass (\ref{hrate2}) is reduced to $M_H\simeq 1/(8\pi)$ 
corresponding to the black hole temperature $T_H\simeq 1$. 
We find that $t_h\simeq 113$, $a_h\simeq 2.84\cdot 10^{28}$ 
and the e-folding factor $N_et_h\simeq 65.5$. 

The reheating process mainly occurs when $t\sim t_h$, where 
$|\Delta\rho_\Lambda(t)|\rightarrow\bar \rho_\Lambda$, $\rho(t)\rightarrow 0$ 
and scaling factor $a(t)$ slowly varies.  The reheating temperature $T_h$ 
can be possibly estimated by $g_s{\pi^2\over15}T_h^4\simeq |\Delta\rho_\Lambda(t_h)|$, 
where $g_s$ stands for the summation over contributions of all relativistic 
particles created in the Hawking process. Since all possible relativistic particles 
are created, the reheating temperature $T_h<T_0\simeq 1$. We leave $T_h$ as
a parameter in this letter. For $T_h\sim O(10^{-2})$, an enormous entropy $S_h= (T_ha_h)^3$ is produced.

These numerical values depend on the $\theta$-value, 
the initial time $t_0$, temperature $T_0$ and scale factor $a_0$.

\vskip0.2cm
\noindent{\it Cosmological constant.}\hskip0.3cm
After reheating to the temperature $T_h$, the nucleation of quantum black holes starts 
again, and the density of energy-gain $\rho^h_\Lambda(t)$ is given by Eq.(\ref{rho})
with the initial time $t_h$ and temperature $T_h$. Because the reheating temperature 
$T_h<T_0\simeq 1$, $\rho^h_\Lambda(t)$ (\ref{rho}) is exponentially suppressed by lower temperature $T_h$. 
This possibly leads to $\rho^h_\Lambda\ll \rho_m\simeq T_h^4$ in Eq.(\ref{requation}). As a result,
the Universe begins the evolution described by the Standard Cosmology:
\begin{equation}
a(t)= a_h (t/t_h)^\alpha,\hskip0.3cm T(t)=T_h (t/t_h)^{-\alpha},
\label{radiative1} 
\end{equation}
with total entropy $S_h=(a_hT_h)^3$. We consider that this is a new era initiated with $t_h,T_h$ and $a_h$,
independently from the inflationary era before the reheating.

The energy-density $\rho^h_\Lambda(t)$ is mainly contributed from quantum black holes nucleated 
in the reheating. Due to the Hawking radiation, the variation of energy-density $\Delta\rho^h_\Lambda(t)$ 
is given by Eq.(\ref{deltarho}) with initial time $t_h$ and temperature $T_h$. 
Analogously to Eq.(\ref{totalrho}), the ``dark-energy'' density is, 
\begin{equation}
\rho_h(t)=\rho^h_\Lambda(t) + \Delta\rho^h_\Lambda(t)<\rho^h_\Lambda(t),
\label{crho} 
\end{equation}
which is related to the cosmological constant $\Lambda=8\pi G\rho_h(t)$. 
Assuming that the mass of quantum black holes has been reduced to the minimal mass $M_H=1/(8\pi)$ 
at the present time $t\gg t_h$, we obtain, 
\begin{equation}
\rho_h(t)= {0.87m_p^4\over 1024\pi^4}\int_{t_h}^td\tau 
\left({1\over T(\tau)}\right)^{(\theta-1)} e^{-{1\over 16\pi T^2(\tau)}},
\label{crho1}
\end{equation} 
from Eqs.(\ref{rho},\ref{deltarho}) and (\ref{crho}). Substituting solutions $a(t)$ and $T(t)$
(\ref{radiative1}) into $\rho_h(t)$, we have,
\begin{eqnarray}
\rho_h(t) &=& {0.87m_p^4\over 1024\pi^4}\int_{t_h}^td\tau 
\left({\tau^\alpha\over T_ht^\alpha_h}\right)^{(\theta-1)} 
e^{-{\tau^{2\alpha}\over 16\pi T^2_ht^{2\alpha}_h}} 
\nonumber\\
&=& {0.87m_p^4\over 1024\pi^4} {(16\pi)^\delta t_h\over 2\alpha }T_h^{1/\alpha}
\Gamma(\delta,z_2,z_1)\label{cosm2}
\end{eqnarray}
where $\delta=[\alpha(\theta-1)+1]/(2\alpha)$, $z_2= t^{2\alpha}/(16\pi T_h^2t^{2\alpha}_h)$ 
and $z_1=1/(16\pi T_h^2)$ in the incomplete Gamma-function,
\begin{equation}
\Gamma(\delta,z_2,z_1) \equiv \int_{z_1}^{z_2}dx x^{\delta-1}e^{-x}
\simeq z_1^{\delta -1}e^{-z_1}.
\label{gamma} 
\end{equation}
The asymptotic representation of Eq.(\ref{gamma}) is for $z_2\gg z_1\gg 1$ and 
$z_2\rightarrow\infty$. Using this asymptotic representation, we 
approximately have a constant energy-density: 
\begin{equation}
\rho_h \simeq {0.87\over 64\pi^3} {m_p^4t_h\over 2\alpha}T_h^{(3-\theta)}
e^{-{1\over16\pi T_h^2}},
\label{cosm3} 
\end{equation}
whose numerical value crucially depends on the reheating temperature $T_h$. 

We find that $\rho_h(t)$ and $\Lambda(t)$
increase, when the Universe is in radiation domination ($t>t_h, T<T_h$), and asymptotically 
approaches the constant (\ref{cosm3}), when the Universe is in matter domination 
($ t\gg t_h, T\ll T_h$).
With $t_h\simeq 113$, $t\sim 10^{61}$, $\alpha=1/2$ in the radiation 
dominant phase, and $\theta=203/45$ for 
the particle content of the Standard Model, we obtain $\rho_h\simeq 6.7 \cdot
10^{-120} m_p^4$ by setting $T_h\simeq 8.45\cdot 10^{-3}$. This is consistent with present observations. 
 
\vskip0.2cm
\noindent{\it Some remarks.}\hskip0.3cm
In vacuum gravitation, the mass-energy of classical matter is zero ($T_{ab}=0$), 
while the zero-point energy of quantum fields is non-zero 
($T^{\rm vec}_{ab}\not =0$). It seems that $T^{\rm vec}_{ab}$ possibly accounts for 
non-vanishing cosmological term in Einstein equation (\ref{equation}). However, 
the positive-energy theorem\cite{yao} actually requires zero mass-energy 
in the Minkowski spacetime. This implies that the zero-point energy of quantum-fields 
in the Minkowski spacetime is not gravitating and should not be related to
the cosmological term in Einstein equation (\ref{equation}). 

We consider that the cosmological term is originated from the vacuum-energy variation
attributed to the unstable quantum fluctuation ($\epsilon \tilde \phi_{ab}$) about
the classical gravitational field ($g_{ab}^S$), i.e., the decay of flat space at finite temperature 
proceeds by nucleating quantum black holes. Based on such a cosmological term, 
we study the evolution of the early Universe and present value of the cosmological constant.  

The density perturbation that leads to the rich content of the present
Universe is possibly originated from quantum fluctuations and Hawking radiation of 
quantum black holes nucleated in the inflationary phase, which will be presented 
in a future work.  
    






\end{document}